\documentclass[amsfonts,showpacs,tightenlines,aps,12pt,floatfix]{revtex4}
\usepackage{bm}
\begin{document}

\title {$Z^*$ resonances: Phenomenology and Models}
\author{Byron K.~Jennings}\email{jennings@triumf.ca}
\affiliation{TRIUMF, 4004 Wesbrook Mall, Vancouver, BC, Canada V6T 2A3}
\author{Kim Maltman}\email{kmaltman@yorku.ca}
\affiliation{Department of Mathematics and Statistics, York University,
4700 Keele St., Toronto ON M3J 1P3 Canada\\
and\\CSSM, University of Adelaide, Adelaide, 5005, Australia}

\date{\today}
\begin{abstract}
We explore the phenomenology of, and models for, the $Z^*$ resonances, the
lowest of which is now well established, and called the $\Theta$. 
We provide an overview of three models which have been
proposed to explain its existence and/or its small width, and 
point out other relevant predictions, and potential problems, 
for each. The relation to what is known about $KN$
scattering, including possible resonance signals in other channels,
is also discussed.
\end{abstract}

\maketitle
\pagebreak

\section{Introduction}

Strangeness $+1$ baryon resonances ($Z^*$'s) 
have been treated with considerable disdain in the past (see,
for example, the comments in the Particle Data Group~\cite{PDG92} for 1992,
the last year they were discussed). Even at that time there 
were candidates for $Z^*$ resonances~\cite{PDG82,arndt85}. 
It is interesting to notice that the paper~\cite{veit85} following 
the latter of these references increased the spin-orbit force by a
factor of three in a cloudy bag model calculation in a desperate attempt to
reduce the need for the $Z^*$ resonances. The results of the 1985 analysis were
largely confirmed in a later analysis~\cite{hyslop92} by the same group and are
roughly consistent with Ref.~\cite{PDG82}. More recently it was shown that
the poles found in the 1992 analysis correspond to peaks in the time delay and
speed plots~\cite{kelkar03}.

Theoretically, multi-quark states were considered long ago in the bag model,
and the masses of $Z^*$ configurations calculated in some detail for the
negative parity sector~\cite{strottman}.  However these states typically suffer
from the presence of a fall-apart mode and are usually associated with poles in
the p-matrix~\cite{jaffe79} rather than with real resonances (poles in the
t-matrix). Even that association has been questioned~\cite{iqbal86}.

In the Skyrme and chiral soliton models of the nucleon, states with exotic
quantum numbers occur naturally through the presence of solutions
corresponding to higher flavor representations. 
In the $SU(2)_F$ case, an early embarrassment
for these models was the prediction of $I=J=5/2, 7/2,\cdots$ states. 
These states arise via projection from the same intrinsic state 
as the ground state. Since the $I=J=5/2,Y=1$ state, in particular,
was not seen, it was assumed to be an artifact of the model. 
(The model is natural in the $N_c=\infty$ limit, but would in general 
require $1/N_c$ corrections in the real world.) 

The $SU(3)_F$ version of the model also predicts a number of 
higher states, these occurring in various exotic multiplets, 
$\overline{\bf 10}_F$, ${\bf 27}_F$, ${\bf 35}_F$, etc..  
The $SU(2)$ $I=J=5/2$ state lies in an $SU(3)_F$ ${\bf 35}_F$ and
the $I=J=7/2$ in an ${\bf 81}_F$. The $Z^*$ resonances with isospin 0, 1 and 2
lie in the 
$\overline{\bf 10}_F$, ${\bf 27}_F$ and ${\bf 35}_F$ representations,
respectively. (In pentaquark models these are the only representations with
strangeness $+1$.) The existence of such
a $\overline{\bf 10}_F$ state was noted long ago~\cite{man84,chemtob85}.
Since such resonances did not correspond to 3-quark states they tended to 
be ignored. This changed with the work of Ref.~\cite{dpp97} and the 
follow-up work of Refs.~\cite{weig98,weig00} where a narrow state in
the 1500-1600 MeV region was predicted (see also
Refs.~\cite{wall03,pras03}). A narrow strangeness $+1$ state was
then found experimentally
in this energy region~\cite{spring8,itep,clas,elsa,adk03,clas2,hermes}
generating a good deal of subsequent theoretical 
discussion~\cite{wall03,pras03,sr03,cpr03,hos03,kl03,jw03,nus03,asw03,ct03,cckn03,schem1,schem2,pentaq,bfk03,cohetal03,ikor03,dpresp03,dp03,poby03,prasnew03,general,prod,dc03,zhu03,mnnsl03,sdo03,cfkk03,sas03}.
It is now
necessary to understand the nature of this state and the implications of its
existence. The existence of a new narrow resonance in a region of the
baryon spectrum thought previously to be reasonably well
understood~\cite{capstick86} raises questions about how good this
understanding actually is. In particular, it raises the possibility
(or, perhaps, likelihood) that states with non-exotic quantum numbers
may be either structurally similar to the recently-observed exotic
(pentaquark) state, or contain significant 
admixtures of such exotic configuration(s).

In this paper we consider the phenomenology of the $Z^*$ resonances, including
the recently discovered $\Theta$. We explore the implications of a number
of models compatible with the existence of a $Z^*$ 
resonance in the region of the $\Theta$ for the as-yet-undetermined
quantum numbers of the $\Theta$, and consider 
other potentially observable $Z^*$ states predicted by those models.

The rest of the paper is organized as follows.
In Sec.~\ref{sec:phen} we summarize the current experimental situation for the
$Z^*$ resonances and related $KN$ scattering results. In
Sec.~\ref{sec:skyrme} we discuss the results from various 
soliton model calculations, with an emphasis on implications 
for possible states beyond the $\Theta (1540)$. In Sec.~\ref{sec:penta} we
discuss models based on an explicit pentaquark structure, involving
interquark interactions mediated by either effective
Goldstone boson, or effective color magnetic, exchange, and compare the
two approaches. In Sec.~\ref{sec:other} we discuss briefly recent
QCD sum rule and lattice explorations of the $I=0,1$ $Z^*$ sectors.
Finally, in Sec.~\ref{sec:conclusion} we draw
conclusions and suggest directions for future work.
 
\section{The $Z^*$ phenomenology}
\label{sec:phen}

In this section we summarize the phenomenology of the $Z^*$ resonances.
We begin with a brief reminder of the basic results from $KN$ scattering.
It is known that both s-wave phase shifts are repulsive at low
energies~\cite{hyslop92}. This implies that the central part of the
$KN$ interaction will produce no $KN$ potential model resonances.
In the p-wave sector, the $P01$ and $P13$ waves are attractive while the
$P03$ and $P11$ waves are repulsive. This suggests a spin-orbit
potential with different signs in the two isospin sectors. These qualitative
features are correctly described in a number of approaches: the cloudy
bag model~\cite{veit85}, the meson exchange picture~\cite{buttgen85,had02},
and both the quark Born term~\cite{barnes94} and resonating group~\cite{sll95} 
approaches to the non-relativistic quark model.

The 1982 version of the ``Review of Particle Properties''~\cite{PDG82} lists
five $Z^*$ resonances: $Z_0(1780)$ (P01), $Z_0(1865)$ (D03), $Z_1(1900)$
(P13), $Z_1(2150)$ and $Z_1(2500)$. The last two have no spin or parity
assignment. An analysis of $K^+p$ scattering in 1985~\cite{arndt85} found
evidence for three states: P13 at 1780 MeV, P11 at 1720 MeV and D15
at 2160 MeV. A more complete analysis~\cite{hyslop92}, 
going to higher energies and
also including $I=0$, found four resonances. Their properties
are listed in lines 2-5 of Table~\ref{table:one}. 
For the last two experimental determinations,
the mass is the pole location, while for the first it 
is the Breit-Wigner peak location. The two should not be 
expected to be identical. Ref.~\cite{hyslop92} also gives
Argand plots which show strong forward looping for the first three of these
states. Ref.~\cite{kelkar03} shows speed and time delay plots which are also
consistent with the resonance interpretation.

\begin{table}
\caption{The experimental $Z^*$ resonances. The mass
range quote for the lowest resonance corresponds
to the range of central values found in the experiments of
Refs.~\protect\cite{spring8,itep,clas,elsa,adk03,clas2,hermes}. 
The limit on the width 
is a conservative one, compatible with the limits
reported in all of Refs.~\protect\cite{nus03,asw03,ct03}. 
The parameters of the higher states are 
from Ref.~\protect\cite{hyslop92}.}\label{table:one}
\begin{tabular}{|lll|}
\hline
Mass & Width & Quantum \\
 (MeV) & (MeV) & Numbers\\
\hline
1528-1555  & $<6$  & $(I,J^P)=(0?,?)$\\
1788  & 340 &   D03\\
1811  & 236 &   P13\\
1831  & 190 &   P01\\
2074  & 503 &   D15\\
\hline
\end{tabular}
\end{table}

In addition to the phase shift analysis, there are 
a number of recent photo-production
experiments~\cite{spring8,clas,elsa,clas2} 
all of which see a narrow resonance at
$m\sim$1540 MeV, with a width consistent with experimental
resolution. Explicitly the results are: $m=1540\pm 10$ MeV,
$\Gamma<25$ MeV~\cite{spring8}; 
$m=1543\pm 5$ MeV, $\Gamma<22$ MeV~\cite{clas}; $m=1540\pm 5$ MeV,
$\Gamma<25$ MeV~\cite{elsa}; and $m=1555\pm 10$ MeV, 
$\Gamma <26$ MeV~\cite{clas2}. Two of the experiments, 
Refs.~\cite{elsa,clas2}, report negative results in their searches
for a $\Theta^{++}$ signal. A narrow signal in the same region
has also been seen in $K^+Xe$ scattering~\cite{itep},
as well as in a recent re-analysis of old $\nu /\bar{\nu}$ bubble chamber 
data~\cite{adk03}. The former finds $m=1539\pm 2$ MeV, 
$\Gamma<9$ MeV~\cite{itep}, the latter 
$m=1533\pm 5$ MeV, $\Gamma <20$ MeV~\cite{adk03}.
The HERMES Collaboration
has also presented evidence for a narrow 
$\Theta^+$ in $eD\rightarrow pK_sX$ at a beam energy
of $27.6$ GeV, with $m=1528\pm 3\pm 2$ MeV, $\Gamma <20$ MeV~\cite{hermes},
and no sign of a $\Theta^{++}$ signal. 
The failure to observe a $\Theta^{++}$ in $\gamma p\rightarrow K^-
K^+p$, in both the SAPHIR~\cite{elsa} and CLAS~\cite{clas2}
experiments, if correct, rules out the proposed isotensor assignment
suggested in Ref.~\cite{cpr03}. An $I=0$ assignment is most natural
in light of these results.  However, an $I=1$ assignment is still
possible since, in $\gamma p\rightarrow \bar{K} KN$, three different
reduced isospin matrix elements appear if $I_\Theta =1$, allowing a
cancellation to occur in the $K^-K^+p$ production amplitude.  The
absence in the HERMES experiment, however, 
is very unnatural for an $I=1$ $\Theta$ in the higher
multiplicity production environment and hence, it seems to us,
strongly favor $I=0$.

In addition to the direct upper limits on the width already noted,
indirect upper limits have been obtained using information from elastic
scattering. Ref.~\cite{nus03} gives $\Gamma<6$ MeV, while
Ref.~\cite{asw03} finds a limit of a few MeV, and prefers a width of
an MeV or smaller. Bounds of $< 1-4$ MeV (from a consideration of
$K^+d$ scattering data) and $<0.9\pm 0.3$ MeV (from a consideration of
the signal in the DIANA experiment~\cite{itep}) have also been
obtained in Ref.~\cite{ct03}.

In settling on the entries for the mass and width of the $\Theta (1540)$ 
in Table~\ref{table:one}, we have taken a conservative approach and
quoted the full range
of central values for the mass obtained in the various experiments,
and the largest of the reported upper bounds. It is likely that
the width is significantly smaller than this upper bound.

The example of the $\Theta$ shows that important information can
be obtained from knowing that no resonance has been seen
in a given energy region in the existing $KN$ database.
Indeed, we must take into account not only the resonances that
have been claimed, but also the absence of any other resonance signals.
The $KN$ phase shift analysis~\cite{hyslop92}
saw no resonances below 1788 MeV. A resonance could have been missed
if it was too narrow (as in the case of the $\Theta(1540)$) or if it 
was very wide. A medium-width resonance
should have been seen, if it exists. Similarly, Refs.~\cite{clas,elsa,clas2} 
would have seen a resonance if it were narrow, at least if its mass were
below $\sim 1800$ MeV. Refs.~\cite{elsa,clas2}, 
in particular, were able to rule out
$I=2$ states in this range as a result of
their increased sensitivity to the $+2$ charge state.

The nominal threshold for the $K\Delta$ channel is 1725 MeV. A state with
a mass significantly below this value can decay only to $KN$
or, with additional phase space suppression, $K\pi N$. We would expect
such a state to have been seen, if it exists. Above $\sim$1725 MeV, 
the opening of the $K\Delta$ channel could make an $I=1$ or $2$ state broad.
The $K^*N$ threshold is at 1830 MeV and gives an additional open channel
for higher mass states with $I=0$ or $1$.

We conclude that below $\sim 1725$ MeV there is 
likely only one $Y=2$ resonance, the 
$\Theta(1540)$. Above $\sim 1725$ MeV,
a resonance may have been missed only if it is very broad and
decays predominately into a channel other that $KN$. 

\section{The Chiral Soliton Model}
\label{sec:skyrme}

Here we review the results obtained from the Skyrme and Chiral soliton 
models. We rely on 
Refs.~\cite{dpp97,weig98,weig00,wall03,pras03,bfk03,ikor03,dp03}.
Our aim is (i) to explore the extent to which
these models make predictions different from those
of the pentaquark models presented in the
next section and (ii) to see what additional experimental information 
would serve to best test the soliton model approach.

It is worth noting
that there has been some recent debate over the validity of
the rigid rotor approximation for the quantization of the
relevant collective modes in the soliton picture, in particular
concerning the relation of this approximation
to the large $N_c$ limit of QCD~\cite{cohetal03,dpresp03,poby03,cohennew}.
Potential problems with the rigid rotor approximation for
$S=+1$ exotic states were noted long ago~\cite{kk90}. More recently, 
the question of whether it is safe to neglect the couplings between 
collective rotational modes and other degrees of freedom 
at large $N_c$ was raised in connection with the observation
that the splitting of exotic from non-exotic baryon states does not go to 
zero as $N_c\rightarrow\infty$~\cite{cohetal03}.
That the rigid rotor approximation is not necessarily exact in
the exotic sector, even in the large $N_c$ limit,
is seen explicitly in the toy model constructed by Pobylitsa~\cite{poby03}.
Cohen~\cite{cohennew} has also provided arguments showing that,
in general, in the exotic sector, the  exotic collective rotational
modes need not decouple from the vibrational modes, even as
$N_c\rightarrow\infty$. Significant vibrational-rotation coupling
was also seen explicitly, for $N_c=3$, in the results of 
Ref.~\cite{weig98}. None of these observations, however, rules
out the possibility that treating the collective rotations
as the dominant degrees of freedom in the low energy part of
the spectrum might be a good phenomenological approximation. 
Vibrational-rotational mixing is , in fact, likely to be most important for 
states with non-exotic quantum numbers, where non-exotic radially
excited configurations and low-lying exotic configurations 
with the same quantum numbers may lie close together.
This expectation is borne out in the 
explicit calculations of Ref.~\cite{weig98}.

There is a consensus among the various soliton model calculations 
that a state with P01, $Y=2$ quantum numbers and a relatively narrow width
should occur in the 1500-1600 MeV mass range. This state lies
in a $\overline{\bf 10}_F$. A relatively narrow
result for the width of this state was reported in Ref.~\cite{dpp97},
though the precise value quoted ($15$ MeV) has been subsequently
questioned~\cite{footnote1}. A corrected version of
the DPP calculation, given by Jaffe~\cite{jaffe04}, yields
instead $\Gamma_\theta \sim 30$ MeV.
The narrow width results in part from a cancellation~\cite{dpp97,weig98} 
between the contributions of operators proportional to parameters
$G_0$ and $G_1$ which are, respectively, leading and next-to-leading 
order in the $1/N_c$ expansion.
It has been shown, however, that in the large $N_c$ counting,
the coefficient of the $1/N_c$ operator matrix element
receives an $O(N_c)$ enhancement relative to the
leading order operator contribution, so the two canceling
contributions are formally of the same order in $N_c$~\cite{prasnew03}.
A width for the $\Theta$ significantly less than the width
of the $\Delta$ is, therefore, quite natural in the soliton 
picture~\cite{prasnew03}. To satisfy the 
$\Gamma_\Theta <6\, (\sim 1$?$)$ MeV experimental bound, 
however, the numerical cancellation has
to be rather close. Thus, though a relatively narrow
($\sim$ few $10$'s of MeV) width is natural,
a {\it very} narrow ($\sim 1$ MeV) width is less so, since it
requires a rather close fine-tuning of
the magnitudes of the parameters $G_0$ and $G_1$.

In addition to the lowest-lying, P01, $Y=2$ state, there should
be other nearby members of the $\overline{\bf 10}_F$ multiplet. 
The first state of interest here is the
P11, $Y=1$ member, having nucleon quantum numbers.

In the original work of Ref.~\cite{dpp97} (DPP), the $\overline{\bf 10}_F$ $N$
state was identified with the $N(1710)$, which identification was used
to fix the average mass of the $\overline{\bf 10}_F$ multiplet. 
(This is no longer true of the most recent update~\cite{dp03};
we will discuss the updated version below.)
The splitting within the $\overline{\bf 10}_F$ was 
determined by the then-current value ($\sim 45$ MeV~\cite{gls}) of the
nucleon sigma term. The $\overline{\bf 10}_F$ assignment for the
$N(1710)$ is subject to several possible objections.
In Ref.~\cite{weig98}, for instance, once $SU(3)_F$ breaking was taken into
account, the $N$ state in this region was found to have sizable
components not only of the $\overline{\bf 10}_F$
configuration, but also of the radially excited ${\bf 8}_F$ state, as
well as states in the ${\bf 27}_F$ and ${\bf 35}_F$ multiplets.
There are also potential problems associated with
the predicted decay widths: the soliton model predicts
the $\Delta\pi$ decay width to be a factor of $2.2$ smaller
than the $N\eta$ decay width for a pure $\overline{\bf 10}_F$ state,
whereas the $N(1710)$ has a large $\Delta\pi$ but small
$N\eta$ branching fraction~\cite{pdg02}. 
(A large relative $N\eta$ branching fraction is also predicted in
the positive parity pentaquark scenario of Ref.~\cite{jw03}.)
In addition, the more complete calculation of 
Refs.~\cite{weig00,wall03} shows two states
close together in this mass range, one coming from the rotation, and
one from the vibration of the soliton. 
(The claim of Ref.~\cite{jw03} that the soliton model does not have a 
nearby ${\bf 8}_F$ with which the $\overline{\bf 10}_F$ can mix
ignores the presence of the vibrational states.
The Roper would be predominately a vibrational state in the
soliton picture.) The vibrational (${\bf 8}_F$) 
and rotational ($\overline{\bf 10}_F$) states mix strongly in
the analysis of Refs.~\cite{weig98,wall03}. This mixing will,
no doubt, have an important effect on the 
predicted branching ratios. The analysis of
Ref.~\cite{batinic95} does, in fact, suggest that there are two 
$N$ states in this region. Two states are probably 
necessary in both the soliton and positive
parity pentaquark models (see the next section for a discussion of the
pentaquarks).

The recent report by the NA49 Collaboration of an exotic 
$\Xi^{--}$ state with $m=1862\pm 2$ MeV and $\Gamma < 18$ MeV ~\cite{na4903},
necessitated a re-fitting of the $SU(3)_F$-breaking parameters
of the original DPP analysis~\cite{dp03}. This re-fitting
turns out to require a different assignment for the $N(1710)$,
which is perhaps welcome in light of the comments above. 
The reason is as follows. The NA49 state
can be naturally interpreted as the $I=3/2$ $S=-2$ $\overline{\bf 10}_F$
partner, $\Xi_{3/2}$, of the $\Theta$. This state was originally
predicted to have a mass of $2070$ MeV by DPP. Assuming the
$\overline{\bf 10}_F$ assignment is correct, and taking 
the $\Theta$ and $\Xi_{3/2}$ masses as input, one
obtains a alternate solution for the symmetry-breaking
parameters of the model. Taking the result of the ChPT analysis for the
light quark mass ratio $2m_s/(m_d+m_u)=24.4\pm 1.5$~\cite{leutwylermq} 
as input, this solution corresponds to a nucleon sigma
term of $77\pm 5$ MeV, somewhat higher than, though not incompatible with, the
more recent experimental determination of Ref.~\cite{pasw01},
$\sigma_N=64\pm 7$ MeV. The interpretation of the NA49 signal as the 
$\overline{\bf 10}_F$ partner of the $\Theta$ is thus
phenomenologically acceptable in the soliton picture. 
With this interpretation, the $\overline{\bf 10}_F$ $N$ state
lies between $1650$ and $1690$ MeV, once one takes into
account mixing with the ground state nucleon, and hence
is no longer to be identified with the $N(1710)$~\cite{dp03}
(though mixing with the radially excited ${\bf 8}_F$ state,
not considered in Ref.~\cite{dp03}, may complicate this
picture). 

It is worth stressing that, in the approach of Refs.~\cite{dpp97,dp03},
once the $\Theta$ and $\Xi_{3/2}$ masses are employed as input,
{\it all} parameters in the model are fully determined. 
Precise predictions then follow for the locations of other exotic baryon 
states. Of particular interest for testing the soliton picture
are those exotic states lying in the next lowest multiplet,
having $J^P=3/2^+$, ${\bf 27}_F$ quantum numbers. 
These states were considered in detail in Ref.~\cite{bfk03},
using the original DPP parametrization for the symmetry-breaking
terms. It turns out that the modified fit necessitated by the
NA49 observation significantly alters the predictions for most
of these exotic states. The results for the masses from
Ref.~\cite{bfk03}, together with the modified results
obtained using the updated parametrization of Ref.~\cite{dp03},
are given in Table~\ref{table27}~\cite{footnotenew}.
We note that (i) the $I=1$ $J^P=3/2^+$ $Z^*$
resonance, denoted $\Theta_1$, lies rather close to the
$\Theta$ ($m_{\Theta_1}-m_\Theta <90$ MeV), (ii) while 
the position of the $\Theta_1$ is only modestly altered by
the updated parametrization, the masses of the remaining ${\bf 27}_F, 3/2^+$
exotics are all significantly lowered, and (iii) with the
updated parametrization, the exotic $J^P=3/2^+$ $I=3/2$ cascade
state, $\Xi_{27}$, is predicted to lie only $46$ MeV above the
analogous $\overline{\bf 10}_F$ $\Xi_{3/2}$ state. The 
lowering of the masses of the ${\bf 27}_F$ exotics will have a
significant impact, through reduced phase space, on the prediction for
the widths of these states~\cite{bfk03}. 
The presence of a ${\bf 27}_F$ $\Xi_{3/2}$
state should be detectable through the existence of a $\Xi^*\pi$ decay
branch, which is $SU(3)_F$-forbidden for a $\overline{\bf 10}_F$
state, but allowed for a ${\bf 27}_F$ state. 

While fitting the $\Theta$ and $\Xi_{3/2}$ masses fixes the
parameters of the DPP version of the soliton approach, one
should bear in mind that $SU(3)_F$ breaking is
treated only to first order in $m_s$ in Refs.~\cite{dpp97,bfk03,dp03}.
It has been argued elsewhere that higher-order-in-$m_s$
corrections may not be negligible~\cite{pras03,wall03}.
The somewhat high value of $\sigma_N$ obtained from the updated
linear-in-$m_s$ fit may also argue for the presence of higher order 
corrections. In order to get a feel for the uncertainties associated
with such differences in implementation of $SU(3)_F$ breaking, 
we compare the results of the updated fit
above to those of Ref.~\cite{wall03}. The latter
were obtained using the same leading-order-in-$N_c$ $O(m_s)$
$SU(3)_F$-breaking operator as in DPP, and one of the two 
next-to-leading-order-in-$N_c$ $O(m_s)$ operators,
but diagonalizing to all orders rather than truncating
at first order in $m_s$. 

\noindent
\begin{center}
\begin{table}
\caption{Predictions for the masses of exotic states in
the $J^P=3/2^+$, ${\bf 27}_F$ multiplet in the DPP implementation of
$SU(3)_F$ breaking in the soliton model picture. The results
of Ref.~\protect\cite{bfk03}, given in the third column, are based
on the original DPP parameter set, the results denoted ``updated''
on the modified set obtained using the $\Theta$ and $\Xi_{3/2}$
masses as input. Masses are given in MeV.}
\begin{tabular}{|llll|}
\hline
State&$(I,S)$&Ref.~\cite{bfk03}&Updated \\
\hline
$\Theta_1$&$(1,1)$&$1595$&$1628$\\
$\Gamma_{27}$&$(2,-1)$&$1904$&$1727$\\
$\Xi_{27}$&$({\frac{3}{2}},-2)$&$2052$&$1908$\\
$\Omega_{27}$&$(1,-3)$&$2200$&$2088$\\
\hline\end{tabular}
\label{table27}
\end{table}
\end{center}

The comparison of the results for the 
nearby exotic states in the two approaches is given in Table~\ref{tablenlo}.
We see that the sensitivity to the treatment of $SU(3)_F$ breaking
is rather modest, the largest discrepancy being $82$ MeV. This
occurs for the case of the $\overline{\bf 10}_F$ $\Xi_{3/2}$ state,
which comes out somewhat low in comparison with experiment 
in the approach of Ref.~\cite{wall03}. One should, however, bear in mind
that the $\Xi_{3/2}$ mass was used as input in fixing the model
parameters in Ref.~\cite{dp03} while the value, $1780$ MeV~\cite{wall03}, 
obtained in Ref.~\cite{wall03} was a prediction, made in advance of the 
NA49 observation. The value of Ref.~\cite{wall03} is rather similar to 
the estimate, $\sim 1750$ MeV, given in pentaquark model of Ref.~\cite{jw03}. 
The agreement between two such apparently different models is 
quite surprising. One should also bear in mind that the
``all-orders'' treatment of Ref.~\cite{wall03} includes only the
higher-order-in-$m_s$ effects generated by diagonalizing the
lead-order-in-$m_s$ operators. Additional effects associated with
higher-order-in-$m_s$ effective operators have been neglected. For the
sake of both (i) verifying that it is possible to reproduce the observed 
$\Xi_{3/2}$ mass in the all-orders-diagonalization approach
and (ii) determining the size of the shifts in the masses
associated solely with the higher order diagonalization corrections,
it would be interesting to add the remaining
next-to-leading-order-in-$N_c$ $SU(3)_F$-breaking operator employed in DPP
to the analysis of Ref.~\cite{wall03}. 
Note that any pentaquark model which predicts the $\Theta$
will also predict exotic $\Xi$ states obtained from the
$\Theta$ by interchanging the strange quark and one species of light quark 
$\bar s \Rightarrow \bar u$, $u \Rightarrow s$ or similarly for the $d$). 
The existence of such states thus does not, by itself, distinguish between 
the soliton and pentaquark pictures. 

\noindent
\begin{center}
\begin{table}
\caption{Results for the masses (in MeV) of the nearby exotic
states lying above the $\Theta$. Results in the column labeled
``Linear'' correspond to the linear-in-$m_s$
treatment with the updated values of the fit parameters 
obtained using the $\Theta$ and $\Xi_{3/2}$ masses as input.
Results in the column labeled ``All Orders'' are those from
Ref.~\protect\cite{wall03}, and
correspond to the all-orders-in-$m_s$ diagonalization
explained in the text.}
\begin{tabular}{|llll|}
\hline
State&$(F,J^P,I,S)$&Linear&All Orders\\
\hline
$\Xi_{3/2}$&$(\overline{10},{\frac{1}{2}}^+,{\frac{3}{2}},-2)$&1862 (fit)
&$1780$\\
$\Theta_1$&$(27,{\frac{3}{2}}^+,1,1)$&1628&$1650$\\
$\Gamma_{27}$&$(27,{\frac{3}{2}}^+,2,-1)$&1727&$1690$\\
$\Xi_{27}$&$(27,{\frac{3}{2}}^+,{\frac{3}{2}},-2)$&1908&$1850$\\
$\Omega_{27}$&$(27,{\frac{3}{2}}^+,1,-3)$&2089&$2020$\\
\hline\end{tabular}
\label{tablenlo}
\end{table}
\end{center}

Let us return to the $\Theta_1$, which is the next lowest lying
$Z^*$ resonance after the $\Theta$ in all of the soliton model analysis.
We have seen that there is only very modest sensitivity to the treatment of
$SU(3)_F$ breaking in the predicted mass of the $\Theta_1$
in the two rigid rotor approaches discussed above. A somewhat
higher mass, $\sim 148$ MeV above the $\Theta$, 
is obtained in the bound state approach to strangeness
in the soliton model~\cite{ikor03}, though one should bear in mind,
as pointed out by the authors of Ref.~\cite{ikor03}, that
somewhat larger-than-expected $SU(3)_F$-breaking modifications
of the parameters of the model are necessary
to accommodate the $\Theta$ in this approach.
An interesting observation made in Ref.~\cite{ikor03} is that,
independent of the details of the implementation of the bound
state approach, a particular linear combination
of the splittings of the $I=1$ ${\bf 27}_F$, $J^P=3/2^+$ and ${\bf 27}_F$, $J^P=1/2^+$
$Z^*$ resonances from the $\Theta$ is determined solely by the pionic
moment of inertia, $I_\pi$, which is very well constrained by the
$\Delta$-$N$ splitting. One thus has the following sum rule relating
the $Z^*$ splittings to the $\Delta$-$N$ splitting:
\begin{equation}
{\frac{2}{3}}\left[ m_{{\bf 27}_F,3/2^+}-m_\Theta\right]
+{\frac{1}{3}}\left[ m_{{\bf 27}_F,1/2^+}-m_\Theta\right]
={\frac{2}{3}}\left[ m_\Delta -m_N\right] =195\ {\rm MeV}.
\end{equation}
Since the ${\bf 27}_F$, $1/2^+$ $Z^*$ state lies significantly above
the ${\bf 27}_F$ $3/2^+$ ($\Theta_1$) state, for any phenomenological
acceptable parametrization of the model, an $I=1$ $\Theta_1$ partner of 
the $\Theta$ appears unavoidable below $\sim 1700$ MeV in this framework.
This feature is thus common to both the bound state and rigid rotor
versions of the chiral soliton model.

Whether or not a relatively low-lying state such as the $\theta_1$
should be observable in existing (or future) experiments depends
on its width. If one employs only the leading-order-in-$N_c$ operator of DPP,
one obtains an estimate $\Gamma_{\Theta_1}\simeq 80$ MeV~\cite{bfk03}.
It seems to us it would be surprising if 
a $S=+1$ state with $\Gamma\sim 80$ MeV had not been seen in either the 
production or scattering experiments discussed in the previous 
section~\cite{footnotewidth}.
Even the higher estimates of $\sim 1650-1690$ MeV for the mass are likely to be
problematic, since the first $Y=2$, P13 state seen in the data is the one at
$1811$ MeV. The spin-isospin excitation energy for the $\Theta$ thus appears 
to be a factor of $\sim$2 or more too small in the soliton
model approach, assuming the earlier experimental results are correct. 
As we will see in the next section, the 
pentaquark picture also predicts a low lying $I=1$ excitation
of the $\Theta$, and hence suffers from the same apparent problem.

The next $Z^*$ state, with $(I,J^P)=(1,1/2^+)$ quantum numbers, is
predicted to lie significantly higher than the corresponding
$(1,3/2^+)$ state (for example, at $2030$ MeV in the rigid 
rotor approximation using the original DPP parameter set~\cite{bfk03},
$1861$ MeV using the updated parameter values, 
and $1830$ MeV in the bound state approach~\cite{ikor03}).
Its width is likely to be rather large~\cite{footnotewidth}.
With a large width, it would become important to consider
corrections to the rigid collective coordinate rotation approximation, which
might significantly affect the prediction for the mass. A broad state could
also easily have escaped detection in the scattering experiments. 

Whether or not the soliton model can quantitatively accommodate the putative 
D03 and D15 resonances is not yet clear. Ref.~\cite{wall03} suggests, based 
on an analogy with non-exotic channels, that the D03 and D15 resonances may be
quadrupole excitations of the lower P01 and P13 resonances. It also (i)
suggests that the $1831$ P01 state may be the radial excitation of the $\Theta$
and (ii) argues for the presence of a low-lying S01 resonance, which has not
been seen.  More detailed calculations of the excited exotic states
in the soliton model
are needed in order to see whether these expectations are actually borne out,
in quantitative terms.

Finally, apropos the proposed isotensor assignment 
for the $\Theta$~\cite{cpr03}, we comment that the lowest-lying
$I=2$ state comes out very high in the soliton model
($\sim 1950$ MeV in Ref.~\cite{wall03} and $2035$ MeV in the bound
state approach~\cite{ikor03}). Rather similar values ($\sim 1980$ MeV)
are obtained in either the Goldstone-boson-exchange 
or color-magnetic-exchange versions of the pentaquark picture.
Thus in all of these approaches the isotensor assignment is strongly
disfavored.

In summary, the soliton model 
accounts fairly well for the observed properties of the
$\Theta$, provided the resonance quantum numbers turn out to be
$(I,J^P)=(0,1/2^+)$. Potential problems for the approach are the need for a
second nucleon state near 1710 MeV, and the location of the $Y=2,\
(I,J^P)=(1,3/2^+)$ state. The exotic $\Xi_{3/2}$ is predicted to
lie somewhat low in one version of the model\cite{wall03} but can 
be accommodated with not-unreasonable parameter values.
Improved experimental data in the energy
region of the problematic states would be useful, as would 
experimental searches for the other predicted exotics, and explicit
calculations for the location of the lowest exotic negative parity states.

If the quantum numbers of the $\Theta(1540)$ are other than 
$(I,J^P)=(0,1/2^+)$ the soliton model is in serious, and probably terminal,
trouble: not only will it have predicted a state that has
not been found, it will have failed to predict a state that has been.

\section{$Z^*$ Resonances as Pentaquarks}
\label{sec:penta}
In order to construct a $Z^*$ state like the $\Theta (1540)$
in the quark model, one requires a configuration with
a minimum of four light ($u,d$) quarks and one $\bar{s}$ quark. 
There is a long history of interest in, and
quark-model-based studies of, channels where such pentaquark
configurations might
occur~\cite{hs78,strottman,veit85,mg86,barnes94,sll95,stancu98,bm99,hr02,had02}.
The discovery of the $\Theta$ has sparked renewed interest in
this~\cite{sr03,kl03,jw03,nus03,cckn03,schem1,jm03,pentaq,schem2,dc03}, as 
well as
other~\cite{hos03,general,zhu03,mnnsl03,sdo03,cfkk03,sas03}, approaches. 
We discuss here
two versions of the quark model approach: one in which the spin-dependent 
$qq$ interactions are generated by effective Goldstone boson exchange between 
constituent quarks~\cite{gr96} (the GB case), 
and one in which they are generated by effective color-magnetic
exchange (the CM case). The bag model and non-relativistic 
constituent quark model represent two different
implementations of the CM approach. We refer to the
interactions in both the GB and CM cases,
collectively, as ``hyperfine'' interactions. 

In the GB case, the effective interaction has the form
\begin{equation}
H_{GB}=\sum_{i<j=1,\cdots ,4}H_{GB}^{ij}=
-C_{GB}\sum_{i<j=1,\cdots ,4,F}
\left[ \vec{\lambda}^F_i\cdot\vec{\lambda}^F_j\right]
\left[ \vec{\sigma}_i\cdot\vec{\sigma}_j\right] f_F(r_{ij})
\label{HGB}\end{equation}
where the sum on $i,j$ runs over the four light
quarks, that on $F$ runs over the octet of
pseudo-Goldstone bosons ($\pi$, $K$, $\eta$), and the form of
$f_F(r_{ij})$ employed in the model may be found in Ref.~\cite{stancu98}.
Note that in Refs.~\cite{sr03,schem2} an approximate, ``schematic'' 
version of $H_{GB}$ was employed, in which the spatial dependence 
of the interaction was omitted. As we will see below, this 
approximation can lead to a significant overestimate of the
hyperfine attraction available in the positive parity sector,
and hence should be treated with some caution. (Ref.~\cite{schem1}
performs a similar schematic treatment of the CM interaction.)
As in Ref.~\cite{stancu98}
we do not include GB-induced interactions between the light quarks and
$\bar{s}$ in the putative pentaquark states in order to
avoid incorporating interactions which would correspond to the 
exchange of Goldstone bosons in the Goldstone boson two-particle
subchannel, $\ell\bar{s}$, ($\ell =u,d$).

In the CM case, the effective interaction has the form
\begin{equation}
H_{CM}=\sum_{i<j=1,\cdots ,5}H_{CM}^{ij}=
-C_{CM}\sum_{i<j=1,\cdots ,5}\left[ \vec{F}^c_i\cdot\vec{F}^c_j\right]
\left[ \vec{\sigma}_i\cdot\vec{\sigma}_j\right] f(r_{ij}) \mu_{ij}
\label{HCM}\end{equation}
where the sum now runs over all pairs (with $5$ labeling the
$\bar{s}$ quark), $\vec{F}^c_i=\vec{\lambda}_i/2$ for $i=1,\cdots ,4$,
and $\vec{F}^c_5=-\vec{\lambda_5}^*/2$. The factor $\mu_{ij}$ is 
defined to be $1$ if $ij$ is a light quark pair. In the $SU(3)_F$
limit $\mu_{i5}\equiv\hat{\mu}$ is also equal to $1$. Phenomenologically,
one requires $\hat{\mu}\simeq 0.6$ in order to account for
$\Lambda$-$\Sigma$ splitting in the model. We will consider
both zero range and finite range versions of $f(r_{ij})$ in
what follows.

Before quoting results for the negative
and positive parity hyperfine expectations in the models,
it is worthwhile pointing out certain generic features associated
with the pentaquark picture.

First note that $Z^*$ resonances with $I=0,1,2$ lie in the $4q$ flavor
multiplets $[f]^{4q}_F=[22]$, $[31]$ and $[4]$, respectively. 
Combining these with the antiquark flavor tableau, $[11]$,
one obtains the $SU(3)_F$ representations
\begin{eqnarray}
&&[22]\otimes [11]={\bf \overline{\bf 10}}\oplus{\bf 8}\nonumber \\
&&[31]\otimes [11]={\bf 27}\oplus{\bf 8}\oplus{\bf 10}\nonumber \\
&&[4]\otimes [11] ={\bf 35}\oplus{\bf 10}\ .
\label{flavordoubling}\end{eqnarray} 
The $Z^*$ states lie in the first of the flavor multiplets 
on the RHS in all cases. These are the
only possible $Z^*$ flavor classifications possible in the
pentaquark picture. In the absence of flavor-dependent
interactions between the antiquark and the quarks (as is
the case for both the GB and CM models outlined above), the multiplets
on the RHS's of Eqs.~(\ref{flavordoubling}) are degenerate
in the $SU(3)_F$ limit. (If one allows configurations
containing $s$ quarks, one also has $[211]\otimes [11]$,
which contains degenerate ${\bf 8}_F$ and ${\bf 1}_F$ pentaquark
configurations.) The non-exotic states of the 
$\overline{\bf 10}_F$, ${\bf 27}_F$ and ${\bf 35}_F$ will thus
mix strongly with the corresponding members of the accompanying
${\bf 8}_F$ and/or ${\bf 10}_F$ multiplets once $SU(3)_F$ breaking
is turned on. As pointed out in Ref.~\cite{jw03}, a natural expectation 
is that this mixing might turn out to be ``ideal'', i.e., to diagonalize the 
$s\bar{s}$ pair number. As noted above, such mixing
between members of exotic and non-exotic flavor multiplets
also occurs in the soliton model, though the multiplets
corresponding to the degenerate sets of the pentaquark scenario
are typically not exactly degenerate in the soliton model case.

Second, note that, in the positive parity sector, where
one presumably has one unit of orbital excitation, one
must combine the 5-quark total spin, $S_T$, with the orbital
$L=1$ to form the total angular momentum, $J$.
In the absence of spin-orbit forces, the 
states of different $J$ formed from the same $S_T$ will
be degenerate. Since, empirically, spin-orbit splittings
in the baryon spectrum are typically rather small, there is
a possibility of relatively nearby spin-orbit partners for any state
in the pentaquark picture. A quantitative estimate has been made
in Ref.~\cite{dc03}, assuming the $\vec{L}\cdot\vec{S}$ 
forces to be generated by effective gluon exchange plus scalar 
confinement. Assuming either of the scenarios of Refs.~\cite{kl03,jw03} for 
the $\Theta$ structure, a splitting of order $10$'s of MeV
between the $\Theta$ and its $J^P=3/2^+$ partner is found,
with a conservative upper bound of $150$ MeV.

Finally we comment on the expected widths of pentaquark $Z^*$ states.
Those states with s-wave $NK$ quantum numbers
lying above $NK$ threshold have fall-apart modes
and hence will not correspond to resonances. In contrast, 
for pentaquark states lying above $KN$ threshold, but with
p- or d-wave $KN$ quantum numbers,
the centrifugal barrier may inhibit the decay
to $KN$. A p-wave $KN$ state at 1540 MeV, in a square
well of hadronic size ($\sim 0.8\ fm$), for example,
has a tunneling width of $\sim 280$ MeV, while a corresponding
d-wave state has a width of only $\sim 20$ MeV. Taking into
account the square of the overlap with $KN$, 
the width of a pentaquark state can be significantly 
smaller than the $KN$ tunneling width. However,
especially in the p-wave case, at $1540$ MeV, one is relatively near
the top of the barrier and so states with
significantly higher mass for which fall-apart p-wave
modes exist are expected to be very broad.
Note, however, that such p-wave
fall-apart modes are available in the positive parity sector
only for those states where the quark spin $S_T$ is $1/2$.
Pentaquark states with $S_T=3/2$ or $5/2$ would require a tensor interaction in
order to decay to $KN$ in a p-wave, and hence need not be
undetectably broad, at least if they 
do not lie too far above the relevant p-wave fall-apart threshold 
($\Delta K$ or $NK^*$ for $S_T=3/2$ and $\Delta K^*$ for $S_T=5/2$).

We now consider the hyperfine expectations for 
possible $Z^*$ states in the GB and CM models,
in both the negative and positive parity sectors. In the 
negative parity case, all five quarks can be put in the lowest spatial
orbital. For positive parity, an orbital excitation is required,
and we consider states for which the orbital symmetry,
classified by the $S_4$ of the four light quarks, is $[31]_L$.

\subsection{Negative Parity $Z^*$ Pentaquarks}
In the $SU(3)_F$ limit, with all five quarks in the lowest 
spatial orbital, one can factor out the common spatial matrix element 
and determine the hyperfine expectations by standard
group-theoretic methods. These results are easily checked by
direct computation. $SU(3)_F$ breaking may also be implemented
by both group-theoretic methods and direct computation,
providing a check on the reliability of the calculations.
In quoting results, we will suppress the constants $C_{GB,CM}$ and
spatial matrix elements throughout.

The results for the GB case are given in Column 2 of Table~\ref{table1}. 
For reference, note that the expectation in the $N$ is $-14$, which 
corresponds to a hyperfine energy of $\sim -420$ MeV with
standard values for the parameters of the model. The results 
for the CM case are given in Columns 3 and 4, which correspond
to the $SU(3)_F$ limit ($\hat{\mu}=1$) and $\hat{\mu}=0.6$, respectively.
For reference, the corresponding expectations in the $N$ and
$K$ are $-2$ and $-4$ for $\hat{\mu}=1$ and $-2$ and $-2.4$
for $\hat{\mu}=0.6$. 
For $I=1$, where configurations with the spin of the four
light quarks, $S_\ell =0,1$ or $2$ are all Pauli allowed, and hence
two possible states with $J=1/2$ or $3/2$ exist, we 
show only the lower of the two eigenvalues.

\begin{table}
\caption{The lowest eigenvalues of $\langle H_{GB}\rangle$ 
and $\langle H_{CM}\rangle$ in those
negative parity $Z^*$ channels allowed by the Pauli principle for the
four light quarks. The results are in units of 
either $C_{GB}$ or $C_{CM}$ times the common spatial matrix
element.}\label{table1}
\begin{tabular}{|lrrr|}
\hline
$(I,J)$&GB\ &CM\ &CM\ \\
&&$\hat{\mu}=1$&$\hat{\mu}=0.6$\\
\hline
$(0,1/2)$&$-9.33$&$-4.67$\ &$-3.33$\ \\
$(0,3/2)$&$-9.33$&$0.33$&\ $-0.33$\ \\
$(1,1/2)$&$-8.00$&$-1.44$\ &$-0.78$\ \\
$(1,3/2)$&$-5.33$&$-3.00$\ &$-1.27$\ \\
$(1,5/2)$&$0.00$&$3.33$\ &$2.80$\ \\
$(2,1/2)$&$2.67$&$7.33$\ &$6.27$\ \\
$(2,3/2)$&$2.67$&$3.33$\ &$3.87$\ \\
\hline
\end{tabular}
\end{table}

From the table, we see that, in all channels, the hyperfine
expectation is either repulsive or significantly less attractive
than in $KN$. This is true for both the GB and CM cases.
Taking the non-hyperfine contributions into account, 
the models both predict the negative parity states to lie
considerably above $KN$ threshold, and also significantly
above $1540$ MeV. One should of course bear in mind that the
model treatments of the one-body energies are subject
to significant uncertainties. In particular, the response of
the vacuum to the presence of an additional quark-antiquark
pair is modeled only rather crudely in the bag model, and
not at all in the GB and CM versions of the constituent quark model.
However, even if one is willing to argue that one-body energies
are significantly overestimated, there is no possible negative parity
channel to which it is possible to consistently assign
the $\Theta$, for reasons which we now explain.

For the CM case, the most attractive 
hyperfine expectation occurs for $(I,J)=(0,1/2)$. 
Such a configuration has a potential 
s-wave $KN$ fall-apart mode and hence must be either bound or non-resonant.
The $\Theta$, which is not bound, can therefore 
not be assigned to the $(0,1/2)$ channel.
This argument can be avoided only for states with $J=3/2$ or $5/2$,
which require a d-wave to decay to $KN$. At $1540$ MeV, such a state
would be very narrow, especially once the overlap with $KN$ was
taken into account. The most attractive by far of these higher-spin channels
is that with $(I,J)=(1,3/2)$. It is, however, impossible to 
assign the $\Theta$ to this channel since, if one did, the more attractive
$(0,1/2)$ channel would lie below $KN$ threshold. This is ruled out
experimentally. Thus, no possible negative parity assignment for the $\Theta$
remains in the CM case.

In the GB case, the most attractive hyperfine interaction occurs for
$(I,J)=(0,1/2)$ and $(0,3/2)$, which are
degenerate. A $(0,3/2)$ state, which requires a d-wave $KN$ decay,
would be narrow if located at $1540$ MeV. The 
accompanying $(0,1/2)$ configuration, with its fall-apart s-wave $KN$ mode
would be non-resonant and not a classification problem. However,
as we will see in the next subsection, the optimal GB hyperfine
attraction in the positive parity sector 
is sufficiently strong that the lowest $Z^*$ state
has positive parity. Since no $Z^*$
resonance is observed below the $\Theta$, it follows that
a negative parity assignment is ruled out also in the GB case.

We conclude that, should the $\Theta$ turn out to 
have negative parity, it will be necessary to abandon 
both the CM and GB models (as
well as the current implementations of the soliton model)
for any future applications in the multiquark sector.

\subsection{Positive Parity $Z^*$ Pentaquarks}
In the positive parity sector, a large number 
of independent Pauli-allowed states exist having $[31]_L$ light quark
orbital symmetry. Since, for a color singlet $4\ell\bar{s}$
state, the light quark color is necessarily
$[211]_C$, the joint spin-isospin-orbital
symmetry of the four light quarks must be $[31]_{ISL}$.
Once one takes into account the coupling of 
the spin of the $\bar{s}$ to $S_\ell$ to form $S_T$, one finds
$4$ such independent $[31]_{ISL}$ states in the
$(I,S_T)=(0,1/2)$ channel, $3$ in the $(0,3/2)$ channel, 
$6$ in the $(1,1/2)$ channel, $5$ in the $(1,3/2)$ channel, 
$2$ each in the $(2,1/2)$ and $(2,3/2)$ channels, and 
$1$ each in the $(0,5/2)$, $(1,5/2)$ and $(2,5/2)$ channels.
The $\ell\bar{s}$ interactions in the CM model couple states
with the same $S_T$, but different $S_\ell$, while
such couplings are absent in the
version of the GB model employed here. While one can simply
construct the full set of states in each channel, compute the
hyperfine expectations, and diagonalize the resulting matrix,
a physical understanding of the origin of the lowest possible
eigenvalues, and a good approximation to the structure of the
corresponding lowest-lying eigenstates, can be obtained from simple physical
arguments based on a consideration of attractive correlations
accessible in the $4\ell\bar{s}$ sector.

In the GB model, the only attractive s-wave $\ell\ell$ correlations
are those with $I=S=0,\ C=\bar{3}$ and $I=S=1,\ C=\bar{3}$. 
In these configurations, the hyperfine pair expectations,
suppressing $C_{GB}$ and the spatial matrix elements, are
$-8$ and $-4/3$, respectively. There are no attractive $\ell\ell\ell$
configurations, apart from the nucleon. Organizing the four light quarks
into two $I=S=0,\ C=\bar{3}$ pairs, thus takes
optimal advantage of the strong hyperfine attraction in that channel.
As pointed out in Refs.~\cite{jw03,nus03}, 
such a two-cluster configuration, coupled to net color $3_C$,
(the ``Jaffe-Wilczek (JW) correlation''),
is forbidden unless the two clusters are in a relative p-wave.
Neglecting interactions between the clusters, as well as
further cross-cluster light-quark antisymmetrization effects
(both suppressed by the relative p-wave between the clusters),
the GB hyperfine expectation for such a configuration 
is $-16$, which is now more attractive than in the $N$.
This configuration is possible only in the $(I,S_T)=(0,1/2)$ channel.
The $(I,S_\ell )=(0,0)$ configuration constructed from two
$I=S=1,\ C=\bar{3}$ pairs, which is also present in the
$(I,S_T)=(0,1/2)$ channel, has a hyperfine attraction of $-8/3$
in the same approximation, and can mix with the JW
correlated state to reduce the hyperfine expectation even further.
We thus expect the hyperfine expectation in the GB model to (i) 
be minimized in the $(I,S_T)=(0,1/2)$ channel, (ii) be less than $-16$,
and (iii) correspond to a state dominated by the JW correlation.
We will see that these expectations are borne out by the
results of the full calculations given below.
In terms of the light quark $S_4\downarrow S_2\times S_2$ substate
labels in the spin, isospin, color and orbital sectors, the JW state,
neglecting cross-cluster antisymmetrization, is
\begin{equation}
\vert JW\rangle =\vert [211]_C AA\rangle\vert [22]_I AA\rangle
\vert [22]_S AA\rangle \vert [31]_L SS\rangle \ .
\end{equation}
With our phase conventions, the $ISC$ overlap of
this state with the $(I,S_T)=(0,1/2)$ $N_{123}K_{45}$ 
configuration is $1/2\sqrt{6}$. The p-wave $KN$ decay width 
will thus be naturally small ($\sim 280(1/24)f^2$ MeV$\simeq 10f^2$ MeV, where
$f$ is a spatial overlap factor) for a state at $1540$ MeV 
dominated by the JW correlation. The next most attractive 
correlation is that involving
one $I=S=0,\ C=\bar{3}$ and one $I=S=1,\ C=\bar{3}$ light quark pair.
This configuration has $(I,S_\ell )=(1,1)$, and produces degenerate
$(I,S_T)=(1,1/2)$ and $(1,3/2)$ configurations
when combined with the $\bar{s}$ quark. The hyperfine expectation is
$-28/3$, before additional mixing is included. 

The situation, though somewhat more complicated, is similar in
the CM case. Here the only attractive $\ell\ell$ correlations are
those with $I=S=0,\ C=\bar{3}$ and $I=0,\ S=1,\ C=6$. The
hyperfine expectations, again suppressing $C_{CM}$ and spatial
matrix elements, are $-2$ and $-1/3$, respectively. A strongly
attractive light quark configuration is again formed by
constructing the JW correlation, whose CM hyperfine expectation
is $-4$. The JW correlation has $(I,S_\ell )=(0,0)$,
and hence is present only in the $(I,S_T)=(0,1/2)$ channel.
The $(I,S_\ell )=(0,1)$ correlation produced by combining
one $I=S=0, \ C=\bar{3}$ and one $I=0,\ S=1, \ C=6$ pair
has a less attractive light quark hyperfine expectation, $-7/3$, and can
also contribute to the $(I,S_T)=(0,1/2)$ channel. As 
first noted by Karliner and Lipkin~\cite{kl03} (KL), 
however, with CM interactions, coupling the $\bar{s}$ spin to the $S=1$ pair
in such a way as to make the total spin of the three-quark correlation
$1/2$ leads to a reversal of the ordering of the hyperfine energies of 
the two correlated light quark states, once the $\ell\bar{s}$ interactions
are taken into account. Indeed, in the $SU(3)_F$
limit, the hyperfine expectation of the JW correlated state, including the
$\bar{s}$, remains unchanged at $-4$, while that of the KL correlated
state is lowered to $-17/3$. For $\hat{\mu}=0.6$
the expectations are $-4$ for the JW state and $-13/3$ for the
KL state. The $\ell\bar{s}$ interactions not only make the KL
correlation lower in energy than the JW correlation, 
but also couple the two correlated
configurations. This effect is especially important for the
$SU(3)_F$-broken case, $\hat{\mu}=0.6$, where the JW and KL
configurations are close to degenerate. The CM
hyperfine matrix in the JW, KL subspace of the $(I,S_T)=(0,1/2)$
channel is, suppressing $C_{CM}$ and the spatial matrix elements,
\begin{equation}
\left(\matrix{ -4&-2\sqrt{3}\hat{\mu}\cr -2\sqrt{3}\hat{\mu}&
-{\frac{7}{3}}-{\frac{10}{3}}\hat{\mu}\cr}\right)\ .
\end{equation}
The lowest eigenvalue is $-8.4$ for $\hat{\mu}=1$
and $-6.2$ for $\hat{\mu}=0.6$, in both cases significantly
lower than expectation for either the JW or KL configuration.
The optimal combination is a roughly equal admixture of 
the JW and KL correlations. The
$ISC$ overlap with $N_{123}N_{45}$ for such a configuration is
$\sim -1/5$ for our phase conventions, again providing a natural
explanation for the narrow width of a state dominated by
such a configuration and lying at $1540$ MeV.
The above results of course neglect cross-cluster interactions,
as well as additional antisymmetrization-induced effects
between the clusters. They also neglect the presence of other
attractive configurations (for example,
$\ell\ell\ell\bar{s}$ with $I=1/2$, color $\bar{3}$ and spin $1$
produced by coupling the $\bar{s}$ to a $\ell\ell\ell$
configuration with $I=1/2$, color ${\bf 8}$ and $S_\ell =1$). 
These are less attractive than the JW and KL configurations, but can
also mix with the above combination of JW and KL states.
As we will see below, the results of the full
calculation (which includes all cross-cluster interactions
and is fully antisymmetrized in the coordinates of the four light quarks)
are in good agreement with the estimates just
given for the optimal hyperfine expectation. This observation
suggests that the optimized combination of KL and JW correlations
dominates the lowest-lying state in the $(I,S_T)=(0,1/2)$
channel.

We now present the results obtained by constructing the full
set of completely antisymmetrized
states allowed in each channel for the light quark 
$[31]_L$ configuration and computing and diagonalizing the
resulting hyperfine matrix. In each case we quote only the
lowest eigenvalue in the channel in question. Details of
the construction and calculations will be presented elsewhere~\cite{jm03next}.

We first comment briefly on the structure of the spatial
matrix elements in the $[31]_L$ sector.
Using the $S_2^{12}\times S_2^{34}$ labeling,
the $[31]_L$ $S_4$ irrep has a basis 
$\{ \vert SS\rangle ,\vert SA\rangle , \vert AS\rangle$\}. 
Writing the hyperfine matrix element between two fully antisymmetrized
states $\vert [n]\rangle$ and $\vert [m]\rangle$ as
\begin{equation}
\langle [n]\vert H_{GB,CM}\vert [m]\rangle
= 6\langle [n]\vert H^{12}_{GB,CM}\vert [m]\rangle
+4 \langle [n]\vert H^{45}_{GB,CM}\vert [m]\rangle\ ,
\end{equation}
one finds that the matrix elements involve the following,
in general non-zero, spatial matrix elements:
$\langle m\vert f(r_{12})\vert m\rangle$ and
$\langle m\vert f(r_{45})\vert m\rangle$, with $m=SS,SA,AS$, and
$\langle SS\vert f(r_{45})\vert SA\rangle$. 
In the ``schematic'' approximation all the diagonal matrix elements
are set equal to $1$ while the off-diagonal $45$ matrix
element is set equal to zero. 
In general, however, the
$\langle AS\vert f(r_{12})\vert AS\rangle$ matrix
element will be suppressed relative to the 
$\langle SS\vert f(r_{12})\vert SS\rangle$ and
$\langle SA\vert f(r_{12})\vert SA\rangle$ matrix elements
(it, in fact, must vanish for zero range interactions). 
In the GB model, the explicit form of the spatial
dependence used in the model results in a suppression
\begin{equation}
{\frac{\langle AS\vert f(r_{12})\vert AS\rangle}
{\langle SS\vert f(r_{12})\vert SS\rangle}}\simeq 0.3
\end{equation}
if one employs a Gaussian wave function with $p$-wave
excitations in the light quark coordinates. This result is
much closer to the zero range than to the schematic
limit. The relations among the other matrix elements are also
not, in general, well-approximated by the schematic
approximation. 

In generating results for the GB case we have
employed the actual spatial dependence employed by the
proponents of the model, but also quote the results
in the ``schematic'' limit for comparison. The results
are given in Table~\ref{table2}.

\begin{table}
\caption{The lowest eigenvalues of $\langle H_{GB}\rangle$ for
positive parity $Z^*$ channels. The results are in units of 
$C_{GB}\, \langle SS\vert f(r_{12})\vert SS\rangle$. The heading ``Schematic''
refers to the schematic treatment (neglect) of the spatial
dependence in $H_{GB}$, the heading ``Realistic'' to the 
use of the explicit spatial dependence described in 
Ref.~\protect\cite{stancu98}.}\label{table2}
\vskip .05in\noindent
\begin{tabular}{|lrr|}
\hline
$(I,J)$&Schematic&Realistic\\
\hline
$(0,1/2)$ &$-28.0$\ \ &$-21.9$\ \ \\
$(0,3/2)$ &$-9.3$\ \ &$-10.5$\ \ \\
$(0,5/2)$ &$4.0$\ \ &$1.7$\ \ \\
$(1,1/2)$ &$-21.3$\ \ &$-17.1$\ \  \\
$(1,3/2)$ &$-21.3$\ \ &$-17.1$\ \ \\
$(1,5/2)$ &$0.0$\ \ &$-0.9$\ \  \\
$(2,1/2)$ &$2.7$\ \ &$0.8$\ \ \\
$(2,3/2)$ &$-8.0$\ \ &$-6.2$\ \  \\
$(2,5/2)$ &$-8.0$\ \ &$-6.2$\ \  \\
\hline
\end{tabular}
\end{table}

For the CM case we quote results for both a ``zero range'' and
``finite range'' treatment of the spatial dependence.
For the zero range case we employ $f(\vec{r}_{ij})=\delta^3(\vec{r}_{ij})$,
while for the finite range case we employ a Gaussian with
width $\sim 1\ fm$. The results are then quoted with
an overall factor of $C_{CM}\, \langle SS\vert f(r_{12})\vert SS\rangle$ 
factored out. The results are given in Table~\ref{table3}.

In both the GB and CM cases, one can use
the schematic limit to test the reliability of the
calculation since, in that limit, the expectations can
again be determined using group-theoretic methods.

\begin{table}
\caption{The lowest eigenvalues of $\langle H_{CM}\rangle$ for
positive parity $Z^*$ channels. ZR and FR denote the zero
range and finite range versions of the CM spatial dependence,
respectively. The results are in units of 
$C_{CM}\, \langle SS\vert f(r_{12})\vert SS\rangle$.}\label{table3}
\begin{tabular}{|lrrrr|}
\hline
$(I,J)$&$\hat{\mu}=1$&$\hat{\mu}=1$&
$\hat{\mu}=0.6$&$\hat{\mu}=0.6$\\
&(ZR)\ &\ (FR)&\ (ZR)&\ (FR)\\
\hline
$(0,1/2)$&$-6.86$\ \ &$-8.26$\ \ &$-5.14$\ \ &$-6.23$ \\
$(0,3/2)$&$-3.82$\ \ &$-3.11$\ \ &$-2.27$\ \ &$-2.01$ \\
$(0,5/2)$&$2.58$\ \ &$3.03$\ \ &$2.01$\ \ &$2.51$ \\
$(1,1/2)$&$-5.60$\ \ &$-7.84$\ \ &$-3.92$\ \ &$-5.81$ \\
$(1,3/2)$&$-2.19$\ \ &$-2.66$\ \ &$-1.46$\ \ &$-1.78$ \\
$(1,5/2)$&$2.58$\ \ &$3.09$\ \ &$2.22$\ \ &$2.61$ \\
$(2,1/2)$&$1.08$\ \ &$-0.49$\ \ &$1.88$\ \ &$0.19$ \\
$(2,3/2)$&$0.09$\ \ &$-2.02$\ \ &$1.17$\ \ &$-0.47$ \\
$(2,5/2)$&$3.33$\ \ &$3.37$\ \ &$3.07$\ \ &$2.91$ \\
\hline
\end{tabular}
\end{table}
\subsection{Comments on the Pentaquark Results}
Although the crudeness of the treatment of vacuum response in
the models prevents the one-body energies, and hence 
also the absolute location of any particular
state, from being reliably predicted,
the relative orderings, as well as the splittings, in the models
are well-determined, and hence amenable to comparison with
experimental data.

In the GB model, the excitation energy to promote one of the
light quarks to a p-wave is $\sim 250$ MeV~\cite{stancu98}.
In the CM case, for correlations of the type expected to dominate the
most favored pentaquark channel, the
excitation energy is expected to be $\sim 210$ MeV~\cite{kl03}.
We find that, in both the GB and CM cases, the increase
in the hyperfine energy in going from the negative to the
positive parity sector is more than enough to compensate
for the orbital excitation energy. The lowest lying
pentaquark state in both models is thus predicted to
have positive parity. In both cases this state
has $I=0$, $S_T=1/2$, and is to be identified with the $\Theta$. 
Thus, depending on the sign of 
any possible spin-orbit force, the quantum numbers of the $\Theta$ 
are predicted to be $I=0$, with $J^P$ either $1/2^+$ or $3/2^+$.
As we will explain below, other phenomenological input strongly favors
the former assignment. The $1/2^+$ state also lies lowest
for the quantitative estimate of the $\vec{L}\cdot\vec{S}$
splitting given in Ref.~\cite{dc03}.
Note that the ``schematic'' approximation
is, in general, rather unreliable. In particular, in the
$(I,S_T)=(0,1/2)$ channel, to which the $\Theta$ must be
assigned, it leads to a $77\%$ overestimate of the 
size of the hyperfine attraction, relative to $KN$, in the GB case.

The next lowest positive parity states in the GB model correspond to the
degenerate pair $(I,S_T)=(1,1/2)$ and $(1,3/2)$, 
predicted to lie at $\sim 1685$ MeV, before spin orbit interactions
are taken into account. Spin orbit splitting will
make either the P11 or P13 state lowest in the first case,
and either the P11 or P15 state lowest in the second. No
resonance has been reported in this region in any of these
channels, though a $KN$ P13 resonance is claimed at $1811$ MeV.
The first spin-isospin excitation of the $\Theta$ thus
appears in the same vicinity as in the soliton 
model, and hence also corresponds
to an excitation energy which is, on current evidence, too small by a factor
of $\sim$2. The other attractive hyperfine state is that
with $(I,S_T)=(0,3/2)$, predicted to lie at $\sim 1855$ MeV.
Since a P01 resonance is seen in this region (at $1831$ MeV),
the GB model naturally accommodates such a state, provided
the spin-orbit couplings favor the low spin state in the
$I=0$ sector. This identification would then simultaneously
require identifying the $\Theta$ with the $J^P=1/2^+$ configuration.
The lowest of the negative parity configurations not having an
s-wave fall-apart mode is predicted to have
quantum numbers $I=0$, $J^P=3/2^-$ and a hyperfine expectation
$\sim 345$ MeV less attractive than the $\Theta$.
Taking into account the orbital excitation energy,
one expects this state to lie at $\sim 1640$ MeV. The
D03 resonance claimed experimentally is located at $1788$ MeV,
so the GB prediction gives a prediction for the
negative parity excitation energy which is 
a factor of $>$2 too small, though the quantum numbers of the
lowest-lying negative parity state are in agreement with the
experimental claim.

In the CM model, the next lowest positive parity state after the $\Theta$
has $(I,S_T)=(1,1/2)$. Depending on the
range of the effective interaction, it lies between $\sim 30$ and
$95$ MeV above the $\Theta$. Although it has a p-wave fall-apart
mode, which is suppressed only by barrier penetration, 
the underlying $KN$ tunneling width (at least for a square well
of hadronic size) does not grow rapidly enough
with energy to make such a state broad unless it lies in
the upper part of this range. For what would
appear to be the more realistic (finite range) version
of the model, therefore, one would expect to have seen 
this state experimentally. The first spin-parity excitation 
of the $\Theta$ (which should correspond to either the
P11 or P13 wave) is thus again predicted to
lie significantly too low in the spectrum, at least for the finite range
version of the CM model. The next positive parity state has 
$(I,S_T)=(0,3/2)$, and is predicted to lie between
$1755$ and $1855$ MeV, depending on the range of the interaction.
If the $J=1/2$ state is favored by the spin-orbit couplings,
then, as in the GB case, this allows the identification
of this state with the claimed P01 resonance at $1831$ MeV,
and forces the choice $J^P=1/2^+$ for the spin-parity of
the $\Theta$ on us. Finally, with the CM interaction, the
$(I,S_T)=(1,3/2)$ state is predicted to lie in the 
range $1815$ to $1870$ MeV. The lowest state, after spin-orbit
coupling, will have either P11 or P15 quantum numbers. No
such state is seen, though it is predicted to lie significantly
above the p-wave $\Delta K$ threshold, and so may be rather
broad. The lowest negative parity configuration without an s-wave
fall-apart mode, as for the GB case, is predicted to have
quantum numbers $(I,J^P)=(0,3/2^-)$. Taking into account the difference
of the hyperfine splitting relative to the $\Theta$ and
the estimated orbital excitation energy~\cite{kl03}, it is expected
to lie $\sim 240$ MeV above the $\Theta$, i.e. near $1780$ MeV.
It is thus natural to identify it with the claimed
D03 resonance at $1788$ MeV.


\section{QCD Sum Rule and Lattice Studies}
\label{sec:other}

In this section we comment briefly on the results obtained
in recent QCD sum rule~\cite{zhu03,mnnsl03,sdo03} and 
lattice~\cite{cfkk03,sas03,liu03} studies. 
Ref.~\cite{sdo03,cfkk03,sas03} all report evidence for a
negative parity assignment for the $\Theta$. The two analysis frameworks
both have a rigorous relation to QCD so, if all approximations
were under control in these studies, the question of the parity
of the $\Theta$ would be settled, and all of the models discussed
above would be ruled out. We show that it not yet possible to
reach such a conclusion.

\subsection{Lattice Studies}

Ref.~\cite{cfkk03} is a quenched study with Wilson gauge and
fermion actions, $L\sim 2\ fm$, and lattice spacing, $a$, varying between
$0.171$ and $0.093\ fm$. Finite size effects were investigated and a linear 
extrapolation in $a$ performed. 
The range of light quark masses, $m_q$, studied corresponds
to $m_\pi$ in the range $\sim 400-650$ MeV. A linear extrapolation
to physical $m_q$ was employed. The $J^P=1/2^\pm$, $I=0,1$ channels were all
investigated. In most cases a single interpolating field was used,
but on the largest lattice, and at the largest quark mass,
variable linear combinations of two interpolating fields were employed.
Studying the resulting $2\times 2$ correlation matrix allowed
a convincing separation of the scattering state from the non-scattering
state, at this $m_q$~\cite{kovacs03}. In the full analysis,
the lowest $Z^*$ resonance was found to occur 
in the $I=0$ $J^P=1/2^-$ channel. One should bear
in mind that the result quoted in Eq. (3.2) of Ref.~\cite{cfkk03},
\begin{equation}
m_{I=0,J^P=1/2^-}=1539\pm 50\  {\rm MeV},\label{cfkkmass}
\end{equation}
corresponds to the smallest value of $a$, and {\it not} to the
result of the continuum extrapolation shown in Figure 4. The latter is
not explicitly quoted but, reading from the figure, would correspond
to roughly
\begin{equation}
m_{I=0,J^P=1/2^-}=1465\pm 115\  {\rm MeV}.\label{cfkkmasscontin}
\end{equation}
The continuum-extrapolated $I=0$, $1/2^+$ state, again from Figure 4,
has a mass~\cite{cfkk03} 
\begin{equation}
m_{I=0,J^P=1/2^+}\simeq 1.9\, m_{I=0,J^P=1/2^-}.
\end{equation}

Ref.~\cite{sas03} is also a quenched study with unimproved
Wilson gauge and fermion action. A single lattice spacing,
$a= 0.068\, fm$, and lattice size, $L\simeq 2.2\, fm$ were employed.
The values of $m_q$ used correspond to $m_\pi$ in the range $600-1000$ MeV
and, again, a linear chiral extrapolation was assumed. A single interpolating
field with $I=0$, coupling to both the $J^P=1/2^\pm$ channels, was
employed and the projection onto individual parities performed. It
is claimed that two plateaus are seen in the effective mass plot,
one corresponding to the $KN$ scattering state, and one to the relevant
$Z^*$ resonance, though this claim has been questioned~\cite{liu03}.
Again, the $1/2^-$ mass is found to be the lower of the two, with
\begin{equation}
m_{I=0,J^P=1/2^-}=1760\pm 90\  {\rm MeV}.\label{sasaki}
\end{equation}
and
\begin{equation}
m_{I=0,J^P=1/2^+}=\left( 1.5\pm 0.1\right)\, m_{I=0,J^P=1/2^-}.
\end{equation}
Note that the chirally extrapolated $N$ and $K$ masses come out
somewhat high in the simulation ($1050\pm 30$ and $520\pm 10$ MeV,
respectively), so if one estimates the $Z^*$ resonance masses
using the chirally-extrapolated values of the ratios to the
threshold mass, $m_N+m_K$, as in Ref.~\cite{cfkk03}, 
the $I=0$, $1/2^-$ mass would be reduced to $1610$ MeV, 
reducing the disagreement with the estimate of Ref.~\cite{cfkk03}.

The third lattice study~\cite{liu03} is again quenched, with only
a single lattice spacing, but employs, 
instead of Wilson fermions, overlap fermions. Much
lighter quark masses are reached than in the other simulations, 
with a minimum pion mass of $m_\pi\simeq 180$ MeV.
Both the $I=0$, $J^P=1/2^\pm$ channels were considered. 
The $KN$ scattering states, and also the ghost state in the
positive parity channel, were all clearly identified, but no signal
for either a positive {\it or} negative parity $\Theta$ was seen.

We comment here that it is almost certainly crucial to push
the simulations down to the low $m_q$ values reported in~\cite{liu03}. 
Given the expected intimate relation between 
pentaquark configurations and states of the excited baryon spectrum, 
a natural place to look for guidance on this issue is 
results of recent lattice studies of the S11 and Roper
resonances~\cite{nstarlattice}. If one takes only those parts
of the results of Ref.~\cite{nstarlattice} corresponding
to the range of $m_q$ employed in either Ref.~\cite{cfkk03}
or Ref.~\cite{sas03}, one finds that
(i) the negative parity $N^*$ lies significantly lower than
the positive parity $N^*$ for all such $m_q$ and
(ii) if one makes a linear extrapolation of the ratios 
$m_{S11}/m_N$ and $m_{Roper}/m_N$ to physical $m_q$,
the resulting ``prediction'' for the physical ratio $m_{Roper}/m_{S11}$
is $\simeq 1.32$ for extrapolation based on 
the range employed in Ref.~\cite{sas03}
and $\simeq 1.25$ for extrapolation based on 
the range employed in Ref.~\cite{cfkk03}, in both cases 
in serious disagreement with experiment~\cite{footnote2}.
The source of the problem turns out to be the use of the linear chiral
extrapolation: the results of the actual simulation 
are far from linear below $m_\pi\simeq 400$ MeV, displaying
a cross-over of the positive and negative parity
levels around $m_\pi\simeq 240$ MeV and producing a central
value for the mass ratio $m_{Roper}/m_{S11}\simeq 0.94$ at
physical $m_q$, in excellent agreement with experiment. Whether or not 
a similar low-$m_q$ behavior is to be expected for the $Z^*$ signals 
found using pentaquark interpolating fields is not known at present.
The $N^*$ results, however, clearly signal the potential dangers
in assuming the validity of a linear extrapolation from the large
quark masses used in the simulations of Refs.~\cite{cfkk03,sas03}.
This indicates to us that reaching a definitive conclusion on the parity of 
the $\Theta$ is not yet possible on the basis of current simulations.

\subsection{QCD sum rule studies}

All three QCD sum rule studies in the literature employ the
Borel transformed dispersive sum rule formulation, with
a single-pole-plus-continuum ansatz for the spectral function
and factorization estimates for condensates of dimension $D=6$ and
higher.

Ref.~\cite{zhu03} considers the $J=1/2$ $I=0,1$ and $2$ channels.
No parity projection is performed, so both negative and positive
parity $Z^*$ states can, in principle, contribute to the
spectral function of the correlators employed. The continuum
threshold parameter, $s_ 0$, is taken to lie between
$3.6$ and $4.4$ GeV$^2$, while the Borel mass, $M$, is varied over
the range $1.5-2.5$ GeV. Since $s_0/M^2$ is less than or
$\simeq 1$ over much of this range, significant continuum contributions
will be present. Masses of $1560$, $1590$ and $1530$ MeV
are quoted for the $I=0,1$ and $2$ $Z^*$ states, respectively, with
errors of order $150$ MeV in all cases.

Ref.~\cite{mnnsl03} considers the $I=0$, $J=1/2$ channel, again
without parity projection. Two interpolating fields, combined with
a variable relative coefficient, are employed. The Borel mass
is varied over the range $2\ {\rm GeV}^2<M^2<3\ {\rm GeV}^2$,
while $s_0=4.0\pm 0.4\ {\rm GeV}^2$, so continuum contributions
will be less significant than in Ref.~\cite{zhu03}. 
A result $m =1550\pm 100$ MeV is quoted for the $\Theta$ mass.

Ref.~\cite{sdo03} (SDO) also considers only the $I=0$, $J=1/2$ channel,
but performs a parity projection to separate the $P=+$ and $P=-$ cases.
A single interpolating field is
employed, with $1\ {\rm GeV}<M<2\ {\rm GeV}$, and 
$3.24\ {\rm GeV}^2<s_0<4.0\ {\rm GeV}^2$. It is argued that
no $Z^*$ signal is seen in the positive parity channel,
while a mass of $\sim 1500$ MeV is quoted in the negative
parity channel.

We now comment on these analyses. The first point of relevance
is that, taking the expressions and OPE parameter values given
in each reference, and varying $s_0$ to look for a stability
window in $M$ for the physical output ($Z^*$ mass), one finds
that no such stability window exists so long as one imposes
the constraint of spectral positivity. The absence of such
a stability window typically signals the existence of problems 
with the approximations used on either the OPE or spectral integral
side of the sum rules (or both). Such problems can arise from
an overly-simple form of the spectral ansatz and/or poor convergence
with $D$ of the integrated OPE series. 
One similarly finds no stability of the putative $Z^*$ mass with respect
to the relative coefficient of the two interpolating
fields used in Ref.~\cite{mnnsl03}, suggesting that
the lowest state has not been successfully separated
from the other spectral contributions. 
In the cases where no parity projection has been performed, the
lack of stability might result from the presence of reasonably
isolated low-lying states in both parity channels. In such a situation,
the form of the spectral ansatz means that the spectral contribution
of the higher of the two low-lying $P=\pm$ states must be
approximated as part of the continuum contribution. Such an approximation
can be rather inaccurate, especially if (as for the pentaquark correlators)
the continuum version of the spectral function is a strongly-increasing
function of $s$. In such a situation, however, the results
of the analysis could still be interpreted, qualitatively, as 
indicating the need for low-lying spectral strength, and hence of
a low-lying pentaquark configuration.

A more serious potential problem 
is the convergence with $D$ of the integrated OPE series.
In general, as the number of elementary fields in the 
composite interpolating operator grows, OPE contributions 
of a given dimension correspond to higher and higher numbers of loops,
and hence receive stronger and stronger numerical loop suppression factors
in their coefficients. At a given Borel mass, $M$, this means that
higher $D$ contributions for pentaquark correlators 
will typically be much more important relative, for example, to 
the well-determined $D=0$ (perturbative) and
$D=4$ contributions than is the case for ordinary meson and
baryon correlators. This is a significant potential problem
since higher $D$ condensates are typically not known phenomenologically
and end up being estimated by the vacuum saturation/factorization
approximation (VSA). This approximation is known to be rather
crude (being in error by a factor of $\sim 1.5-4$ 
for the $D=6$ contributions to various combinations
of vector and axial vector correlators for which
reliable determinations from data exist~\cite{otherd6,cdgm}), 
and hence can be the
source of significant theoretical systematic errors if
higher $D$ contributions are dominant. One can, of course,
in principle, simply go to larger $M$ in order to further
suppress higher $D$ contributions relative to well-known
low-dimension ones, but doing so without simultaneously
increasing $s_0$ leads to a spectral integral dominated by
continuum contributions, and hence to large errors
on any extracted resonance parameters. Increasing $s_0$, however,
is typically not an option since one requires a
realistic ansatz for the spectral function in the
region below $s_0$ and, if $s_0$ is large, that region of
the spectrum can no longer be sensibly approximated by a single
narrow resonance contribution. 

In discussing the question of the convergence with $D$ of the OPE
series, we concentrate on the SDO analysis
since the projection onto the separate parity channels
makes it more likely that the single-pole-plus-continuum
ansatz can be safely employed. After transferring ``continuum'' spectral
contributions to OPE side, the SDO sum rules 
are of the form 
\begin{equation}
\vert \lambda_\pm\vert^2 e^{-m_\pm ^2/M^2}\, =\, {\frac{1}{M^{12}}}\left[
\sum_{D=2n}c_D{\frac{\langle O_D\rangle}{M^D}}
 \pm \sum_{D=2n+1}c_D{\frac{\langle O_D\rangle}{M^{D+1}}}\right],
\label{sumrule}\end{equation}
where $m_\pm$ is the mass of the $J^P=1/2^\pm$ $Z^*$ state, 
$\lambda_\pm$ is the strength of its coupling to the interpolating
field and the $c_D$ depend on $s_0/M^2$. A similar relation holds
for the derivative with respect to $1/M^2$. The ratio of these
two expressions, from which $\vert \lambda_\pm\vert^2$ cancels,
is used to determine $m_\pm$. In SDO the sums on
the RHS include terms up to and including $D=6$. We will now argue
that, at the scales employed in the analysis, large contributions
of $D>6$ are {\it necessarily} present, so that the conclusions
based on including terms only up to $D=6$ are not reliable.

To see this, recall that spectral positivity requires that
the coefficient of the exponential on the LHS of Eq.~(\ref{sumrule})
is positive. If the spectral ansatz is sensible (a necessary
condition for the extracted resonance parameters to have
physical meaning), a negative value for the truncated sum on
the OPE side of the equation {\it necessarily} implies that 
numerically non-negligible positive OPE contributions
have been neglected in the truncation employed. In Table~\ref{sumruletable}
we list the contributions to the OPE side of Eq.~(\ref{sumrule})
for $M=1.5$ GeV (the midpoint of the SDO range)
and all $s_0$ employed by SDO. We see that (i) the $D=5$ contribution
is dominant and (ii) the truncated OPE sum is negative for
$s_0=2.56$ and $3.24$ GeV$^2$. The former observation shows
that the series of the dominant chirally odd (odd dimension)
contributions shows no sign of convergence for any of the $s_0$
considered while the latter demonstrates unambiguously that
$D>6$ OPE contributions cannot be neglected in the positive
parity channel. Since those same contributions enter the
negative parity sum rule, with either the same or opposite
sign, the truncated series for the negative parity case is
also shown to be unreliable. Thus, unfortunately, at those
scales where one can hope to make a sensible ansatz for the
$s<s_0$ part of the spectral function, the convergence of the
OPE is too slow with $D$ to allow a determination of 
the separate negative and positive
parity $Z^*$ masses using the Borel sum rule technique.

\noindent
\begin{center}
\begin{table}
\caption{Contributions to the OPE side of the sum rule of 
Eq.~(\protect\ref{sumrule})
for those terms ($D\leq 6$) included in the analysis of 
Ref.~\protect\cite{sdo03}.
The upper sign for the chirally odd (odd dimension) contributions
corresponds to the positive parity case, the lower sign to the
negative parity case.}
\label{sumruletable}
\begin{tabular}{|llll|}
\hline
&&$s_0$ (GeV)&\\
\hline
$D$&\ \ 2.56&\ \ 3.24&\ \ 4.0 \\
\hline
0&$\ \ .00016$&$\ \ .00052$&$\ \ .00138$ \\
1&$\pm .00009$&$\pm .00027$&$\pm .00065$\\
3&$\pm .00149$&$\pm .00339$&$\pm .00673$\\
4&$\ \ .00040$&$\ \ .00083$&$\ \ .00149$\\
5&$\mp .00315$&$\mp .00576$&$\mp .00943$\\
6&$\ \ .00029$&$\ \ .00047$&$\ \ .00070$\\
\hline\end{tabular}
\end{table}
\end{center}

One might consider trying to redo the sum rule analysis in the
finite energy sum rule (FESR) framework where, through the choice
of the weight function, one has control
over the dimensions of the terms in the OPE which contribute
(up to corrections suppressed by additional powers of $\alpha_s$). 
It is known that the ``pinch-weighted'' version of such FESR's are very well
satisfied at the scales in question, in channels where they have
been tested~\cite{cdgm,kmfesr}. However, with the higher
$D$ chirally odd contributions being numerically dominant
it is almost certainly the case that the integrated
contributions associated with the radiative corrections
in the Wilson coefficients for those higher $D$ terms will 
not be numerically negligible,
even if one has chosen the weight in such a way that the
leading contribution integrates to zero.
It thus appears to us unlikely that the results of such
a FESR analysis would be free of sizable theoretical systematic
uncertainties.

In view of the above comments, we conclude that it is not possible
to fix the parity of the $\Theta$ through the arguments which
currently exist in the literature based on QCD sum rules.

\section{Summary and Discussion}
\label{sec:conclusion}
Although there exist quantitative differences among the three models
discussed above, we have seen that their predictions for the spectrum 
of $Z^*$ resonances are, somewhat surprisingly, 
in qualitative agreement. Specifically
(i) the lowest lying $Z^*$ state is predicted to have positive, not
negative, parity; (ii) the most natural quantum number assignment
for this state is $I=0$, $J^P=1/2^+$; (iii) the first positive 
parity excitation of the $\Theta$ is predicted
to lie significantly lower than indicated by
current experimental data; and (iv) the lowest-lying $I=2$ $Z^*$
excitation is predicted to occur rather high in
the spectrum, near $2$ GeV. 

Although, the arguments presented so far do
not rule out a $J^P=3/2^+$ assignment for the $\Theta$ in
either the GB or CM versions of the pentaquark picture, such an assignment
would create problems in accounting for the claimed P01 state
at $1831$ MeV. An even more compelling argument in favor of
the $J^P=1/2^+$ assignment is given below. Quantum numbers other than
$(I,J^P)=(0,1/2^+)$ for the $\Theta$ would thus represent
a terminal problem for all three models. 

We re-iterate that existing lattice and QCD sum rule analyses do not
yet provide a reliable framework for establishing the parity of the $\Theta$.
In the case of the lattice simulations, a crucial improvement
will be future work at light $m_q$. This is necessary in
order to avoid relying on a linear extrapolation of results obtained
using $m_q$ values for which the analogous extrapolation is known to be
unreliable in the $N^*$ sector.
In the case of the sum rule analyses, the issue is the
convergence in $D$ of the integrated OPE contributions. We have
explained why we are pessimistic about the possibility of improving
the current situation and obtaining 
a reliable sum rule analysis in the $\Theta$ channel.

Even if the quantum numbers of the $\Theta$ do turn out to be
correctly predicted, a significant potential problem, common to
all three models, is the small predicted excitation
energy for the first spin-isospin excitation.
That an $I=1$ $Z^*$ excitation
is expected no more than $150$ MeV above the $\theta$
in all versions of the chiral soliton approach, as well as in
both versions of the pentaquark approach, represents a suprising
commonality between the models. However,
it appears to us unlikely that such a low-lying state would have escaped
detection in both the $KN$ scattering experiments and 
recent photoproduction experiments, especially since it
is not expected to be particularly broad in either the soliton or
pentaquark pictures. It would, of course, be highly desirable
to perform a dedicated search for $Z^*$ states lying above
the $\Theta$, not only to verify that such a low-lying state
has not, in fact, been missed, but also to expand our
empirical knowledge of the $Z^*$ spectrum.
It would also be useful to have predictions in the soliton picture
for the locations of any negative parity $Z^*$ states.

A way in which the soliton and pentaquark predictions {\it may} differ
lies in the potential existence of spin-orbit
partners of the pentaquark states. However, in those cases
where the pentaquark decay is inhibited only by the p-wave centrifugal
barrier, the spin-orbit splitting may push the partner states
up sufficiently far that either the width becomes very
large or the higher state no longer resonates. Whether such
additional states should actually show up in the spectrum or not
is thus likely dependent on specific details, rather than the
basic qualitative features, of the underlying models.
An understanding of the source of the spin-orbit splitting
is probably required to make any progress on this question.

The soliton and pentaquark approaches have many similarities outside the $Z^*$
sector as well. Both predict an exotic $I=3/2$ $\Xi$ state lying in the same
$\overline{\bf 10}_F$ as the $\Theta$. Both approaches also require more than
one $N$ state in the vicinity of the PDG $N(1710)$.  In the case of the 
soliton model, the lowest vibrational excitation is expected to dominate the
Roper. Mixing between another vibrational excitation and the $\overline{\bf
10}_F$ rotational excitation is likely required in order to account for the
dominantly non-strange decay modes of the $N(1710)$.  In the pentaquark
picture, as pointed out by Jaffe and Wilczek~\cite{jw03}, one expects (i)
degenerate ${\bf 8}_F$ and $\overline{\bf 10}_F$ $N$ states in the $SU(3)_F$
limit, and (ii) (probably) ideal mixing of these states after $SU(3)_F$ 
breaking is turned
on. The combination with no hidden strangeness should then lie below the
$\Theta$, and is a natural candidate to dominate the Roper (which comes out
consistently high in $3q$ quark model treatments), while the orthogonal
combination should have hidden strangeness and lie a similar distance {\it
above} the $\Theta$. This second state should, however, as in the 
soliton model
case, decay dominantly to states containing strange particles. Another $N$
state with these properties is thus required in the vicinity of the
$N(1710)$. The $N(1710)$ might then be dominated by a three-light-quark radial
excitation configuration. In either picture one sees that the positive parity
excited baryon spectrum becomes considerably more complicated than previously
thought, once one takes the existence of the $\Theta$ into account.

We would like to stress that the existence of flavor multiplets degenerate in
the $SU(3)_F$ limit is a very general feature, common to {\it any} version of
the pentaquark picture having effective quark-antiquark interactions
independent of flavor. The existence of the $\Theta$, combined with the
observation that the non-strange, ideally mixed combination of the ${\bf 8}_F$
and $\overline{\bf 10}_F$ must lie below the $\Theta$ makes it inevitable that
the pentaquark configuration will play a major role in the structure of the
Roper. In fact, since there are no states significantly below 1540 MeV except
for the Roper and ground state nucleon, both of which have $J^P=1/2^+$, this
argument, in combination with the $I=0$ classification, seems to us to leave no
option other than a $J^P=1/2^+$ assignment for the $\Theta$ in the pentaquark
picture. In addition, with typical non-strange-strange splittings, one would
expect the non-strange state to lie somewhat below the actual Roper (a feature
seen also in the soliton model~\cite{wall03}). To move it up to the observed
location would then require some mixing with a lower state, i.e., the ground
state $N$. This then implies that some level of pentaquark admixture {\it must}
be present in the $N(939)$. A mixing of the $\overline{\bf 10}_F$ into the
nucleon is also seen in the soliton model (see, e.g., the results quoted in
Ref.~\cite{bfk03}).  Similar arguments will hold for the $\Lambda$,
$\Sigma$ and $\Xi$ channels (and even the $\Delta$ channel if we take the
degenerate ${\bf 27}_F\oplus{\bf 8}_F\oplus{\bf 10}_F$ 
excited states into account).
Again we expect the first positive parity excited states to have large
pentaquark components, and the ground states to have at least some level of
pentaquark admixture. The existence of the $\Theta$ seems to us to leave no
way out of these conclusions. This implies that the quark model treatment of
the positive parity baryons must be revisited; certainly for the excited states
and probably also for the ground states.

\begin{acknowledgments}
We thank the Natural Sciences and Engineering Research Council of Canada for
financial assistance, and Jean-Marc Sparenberg for providing us with
results for the tunneling widths of the p-wave and d-wave $KN$ configurations.
\end{acknowledgments}


\begin{thebibliography}{99}
\bibitem{footnote1}The apparent numerical error in the
computation of $\Gamma_\theta$ in Ref.~\cite{dpp97} was first
pointed out in Ref.~\cite{weig98}. A slightly different
calculation of the width in the latter reference obtains 
instead a value of $\sim$40 MeV.
After the preparation of the present manuscript,
a corrected version of the DPP calculation,
employing both the DPP formalism and the central DPP value for the 
ratio $G_1/G_0$, was provided by Jaffe~\cite{jaffe04}.
The result of this calculation is $\Gamma_\theta \simeq 30$ MeV, 
as quoted already in the text.
\end{thebibliography}
\end{document}